\documentclass[aps,prb,twocolumn,superscriptaddress,showpacs,showkeys]{revtex4-2}

\usepackage{color}
\usepackage{graphicx}% Include figure files
\usepackage{dcolumn}% Align table columns on decimal point
\usepackage{bm}% bold math
\usepackage{float}
\usepackage[mathlines]{lineno}% Enable numbering of text and display math
\usepackage{tabularx}
\usepackage[colorlinks,linkcolor=blue,anchorcolor=blue,citecolor=blue,urlcolor=blue,hyperindex,CJKbookmarks,dvipdfm]{hyperref}
\usepackage{footnote}
\usepackage{amsmath}
\usepackage{txfonts}
\usepackage{mathptmx}
\usepackage{ragged2e}
\usepackage{booktabs,makecell, multirow, tabularx}
\usepackage{multirow}
\usepackage{braket}

\begin{document}

\title{Hund electronic correlation in La$_3$Ni$_2$O$_7$ under high pressure}

\author{Zhenfeng Ouyang}\affiliation{Department of Physics and Beijing Key Laboratory of Opto-electronic Functional Materials $\&$ Micro-nano Devices, Renmin University of China, Beijing 100872, China}\affiliation{Key Laboratory of Quantum State Construction and Manipulation (Ministry of Education), Renmin University of China, Beijing 100872, China}
\author{Jia-Ming Wang}\affiliation{Department of Physics and Beijing Key Laboratory of Opto-electronic Functional Materials $\&$ Micro-nano Devices, Renmin University of China, Beijing 100872, China}\affiliation{Key Laboratory of Quantum State Construction and Manipulation (Ministry of Education), Renmin University of China, Beijing 100872, China}
\author{Jing-Xuan Wang}\affiliation{Department of Physics and Beijing Key Laboratory of Opto-electronic Functional Materials $\&$ Micro-nano Devices, Renmin University of China, Beijing 100872, China}\affiliation{Key Laboratory of Quantum State Construction and Manipulation (Ministry of Education), Renmin University of China, Beijing 100872, China}
\author{Rong-Qiang He}\email{rqhe@ruc.edu.cn}\affiliation{Department of Physics and Beijing Key Laboratory of Opto-electronic Functional Materials $\&$ Micro-nano Devices, Renmin University of China, Beijing 100872, China}\affiliation{Key Laboratory of Quantum State Construction and Manipulation (Ministry of Education), Renmin University of China, Beijing 100872, China}
\author{Li Huang}\email{lihuang.dmft@gmail.com}\affiliation{Science and Technology on Surface Physics and Chemistry Laboratory, P.O. Box 9-35, Jiangyou 621908, China}
\author{Zhong-Yi Lu}\email{zlu@ruc.edu.cn}\affiliation{Department of Physics and Beijing Key Laboratory of Opto-electronic Functional Materials $\&$ Micro-nano Devices, Renmin University of China, Beijing 100872, China}\affiliation{Key Laboratory of Quantum State Construction and Manipulation (Ministry of Education), Renmin University of China, Beijing 100872, China}

\date{\today}

\begin{abstract}
By means of density functional theory plus dynamical mean-field theory (DFT+DMFT), we investigate the correlated electronic structures of La$_3$Ni$_2$O$_7$ under high pressure. Our calculations show that La$_3$Ni$_2$O$_7$ is a multi-orbital Hund metal. Both the 3$d_{z^2}$ and 3$d_{x^2 - y^2}$ orbitals of Ni are close to be half filled and contribute the bands across the Fermi level. Band renormalization and orbital selective electronic correlation are observed. Through imaginary-time correlation functions, the discovery of high-spin configuration, spin-frozen phase, and spin-orbital separation shows that the system is in a frozen moment phase at high temperatures above 290 K and is a Fermi liquid at low temperatures, which is further comfirmed by the calculated spin, orbital, and charge susceptibilities under high temperatures. Our study uncovers Hundness in La$_3$Ni$_2$O$_7$ under high pressure.
\end{abstract}

\pacs{}

\maketitle

\section{Introduction}
It is well-known that a proper description of the electronic correlation effect is the key to understanding unconventional high-temperature (high-$T_c$) superconductors. Since the discovery of cuprate superconductors~\cite{Bednorz-ZPBM64}, numerous scientists have tried to reveal the mechanism of high-$T_c$ superconduvtivity in the past decades. Although the mechanism of superconductivity in cuprates still remains a puzzle, there are some essential consensuses for cuprates. The parent material of cuprates is considered to be a charge-transfer insulator with strong electronic correlation. Cu atoms have 3$d^9$ configuration, and due to the effect of crystal field, 3$d_{x^ 2-y^2}$ orbitals are half filled. By applying hole doping, Cu-3$d_{x^2-y^2}$ orbitals hybridize with O-2$p_x$/$p_y$ in-plane orbitals, forming the effective singlet band with zero spin moment which is known as the Zhang-Rice singlets~\cite{Zhang-PRB37}. And the superconducting pairing mechanism may be related to the antiferromagnetic spin fluctuations.

Many efforts have been attempted to search other transition metal compounds analogous to cuprates. As a neighbor of Cu in the periodic table, Ni element was once expected as a potential candidate to reproduce the high-$T_c$ superconductivity. However, no superconductivity trace was observed in hole-doped nickelates La$_{2-x}$Sr$_x$NiO$_4$~\cite{Chen-PRL71,Cheong-PRB49}, which has the same structure as La$_2$CuO$_4$. Until 2019, the superconductivity was finally discovered in hole-doped infinite-layer RNiO$_2$ thin films (R = La, Nd). With 20$\%$ Sr doing, R$_{0.8}$Sr$_{0.2}$NiO$_2$ shows superconductivity with $T_c$ of 9-15 K~\cite{Li-nature572}. Although the superconducting $T_c$ is much lower than those of the cuprates, a series of phenomena in nickelates such as self-doping effect~\cite{Gu-Innovation3}, absence of long range magnetic order~\cite{Hayward-JACS121,Hayward-SoilidStateSci5} and multiorbital Hund's physics~\cite{Wang-PRB102} provide some new understandings for unconventional superconductivity.

Very recently, the superconductivity was discovered in Ruddlesden-Popper bilayer perovskite nickelate La$_3$Ni$_2$O$_7$ with a maximum $T_c$ of about 80 K under pressure~\cite{Sun-nature}. This is undoubtedly a remarkable breakthrough in the field of high-$T_c$ superconductivity since the discovery of cuprates and iron-based superconductors. The XRD patterns show that La$_3$Ni$_2$O$_7$ possesses the orthorhombic $Amam$ space group from 1.6 to 10 GPa and transforms to $Fmmm$ space group above 15 GPa. And the DFT calculations show that applying pressure could metallize the $\sigma$-bond formed by Ni and the apical O, which may be a crucial clue for superconductivity. For the electronic structure of high pressure phase La$_3$Ni$_2$O$_7$, many theoretical calculations reveal that $e_g$ orbitals of Ni and 2$p$ orbitals of O contribute the bands around the Fermi level~\cite{Luo-arXiv, Lechermann-arXiv, Zhang-arXiv, Sakakibara-arXiv, Gu-arXiv, Christiansson-arXiv, Shilenko-arXiv, Cao-arXiv, Chen-arXiv}. La$_3$Ni$_2$O$_7$ is experimentally considered to be a paramagnetic metal~\cite{Liu-SCPMA66}, while a large interorbital hopping between 3$d_{z^2}$ and 3$d_{x^2 - y^2}$ via in-plane 2$p_x$ or 2$p_y$ orbitals of O is found, which indicates a possible in-plane FM tendency in La$_3$Ni$_2$O$_7$~\cite{Zhang-arXiv}. In addition, a bilayer two-orbital model has been constructed by Luo \emph{et al.}, and the spin susceptibility studied by an RPA method shows that the magnetic signal mostly comes from the $d_{z^2}$ orbitals~\cite{Luo-arXiv}. Moreover, the superconductivity in La$_3$Ni$_2$O$_7$ has been extensively studied. A possible $s_\pm$ wave superconducting pairing in La$_3$Ni$_2$O$_7$ has been proposed~\cite{Yang-arXiv}. The model calculations by functional renormalization group method~\cite{Gu-arXiv} and cluster dynamical mean field theory~\cite{Tian-arXiv} both show the pairing tendency of $s_\pm$ wave. Besides, a minimal effective strong coupling model based on local inter-layer spin singlets of Ni-3$d_{z^2}$ electrons has been proposed, which highlights the importance of the bilayer structure of  La$_3$Ni$_2$O$_7$~\cite{Yang-arXiv2}. As a crucial factor to understanding superconductivity, the electronic correlation in La$_3$Ni$_2$O$_7$ is still worth in-depth discussing to figure out the origin of electronic correlation in La$_3$Ni$_2$O$_7$.

In this work, we investigated the electronic structures and correlation of La$_3$Ni$_2$O$_7$ by means of density functional theory plus dynamical mean-field theory (DFT+DMFT). We performed the calculations under different finite temperatures. Our results show that La$_3$Ni$_2$O$_7$ is a multi-orbital Hund metal with orbital selective electronic correlation. The high spin states and spin-orbital separation revealed by our calculations suggest Hundness rather than Mottness in the system. And the calculated spin, orbital, and charge susceptibilities under high temperatures confirm that the electronic correlation in La$_3$Ni$_2$O$_7$ is attributed to Hund's physics. Our work discovers Hundness in La$_3$Ni$_2$O$_7$ under high pressure and provides an understanding to the origin of its electronic correlation.

\section{Method}
We performed DFT+DMFT calculations to study the electronic correlation of La$_3$Ni$_2$O$_7$ under different finite temperatures. The DFT parts were done by WIEN2K code and the full-potential linearized augmented plane-wave (FP-LAPW) method was implemented~\cite{Blaha-JCP152}. The optimized unit cell including 12 atoms was used. The cutoff parameter was $R_{\text{MT}}K_{\text{max}} =$ 7.0. The muffin-tin radii for La, Ni, and O atoms were fixed to be 2.29, 1.86 and 1.60 bohr, respectively. The generalized gradient approximation (GGA) with Perdew-Burke-Ernzerhof (PBE) functional was chosen as the exchange and correlation potential~\cite{Perdew-PRL77}. The $k$-point mesh for the Brillouin zone integration was 12 $\times$ 12 $\times$ 12. The EDMFTF software package was used to perform the charge fully self-consistent DFT+DMFT calculations~\cite{Haule-PRB81}. A good convergence can be achieved within 50 DFT+DMFT cycles. Each DFT+DMFT cycle contained one-shot DMFT calculation and maximum 100 DFT iterrations. The convergence criteria for charge and total energy were $10^{-7}$ eV and $10^{-7}$ Ry, respectively. We enforced the system to be paramagnetic. Only 3$d_{z^2}$ and 3$d_{x^2-y^2}$ orbitals of Ni were treated as correlated, considering that $t_{2g}$ orbitals all are fully filled. And a single impurity problem was constructed due to all Ni atoms being equivalent. We chosed the Coulomb interaction parameter $U =$ 5.0 eV, and the Hund's exchange parameter $J_H =$ 1.0 eV, which are typical values for nickelates~\cite{Sakakibara-PRL125,Leonov-PRB101}. And the density-density form of the Coulomb repulsion was used. We used the projectors with an energy window from $-$10 to 10 eV with respect to the Fermi level to construct the correlated orbitals. The impurity problem was solved by hybridization expansion continuous-time quantum impurity solver (CT-HYB)~\cite{Haule-PRB75} with exact double-counting scheme for the self-energy function, which was developed by Haule~\cite{Haule-PRL115}. The real frequency self-energy function was obtained by analytical continuation with the maximum entropy~\cite{Jarrell-PR269}. Then it was used to calculate the momentum-resolved spectral function and the other related physical quantities.

\section{Results and discussion}

\subsection{Band renormalization}
\begin{figure}[htb]
\centering
\includegraphics[width=8.6cm]{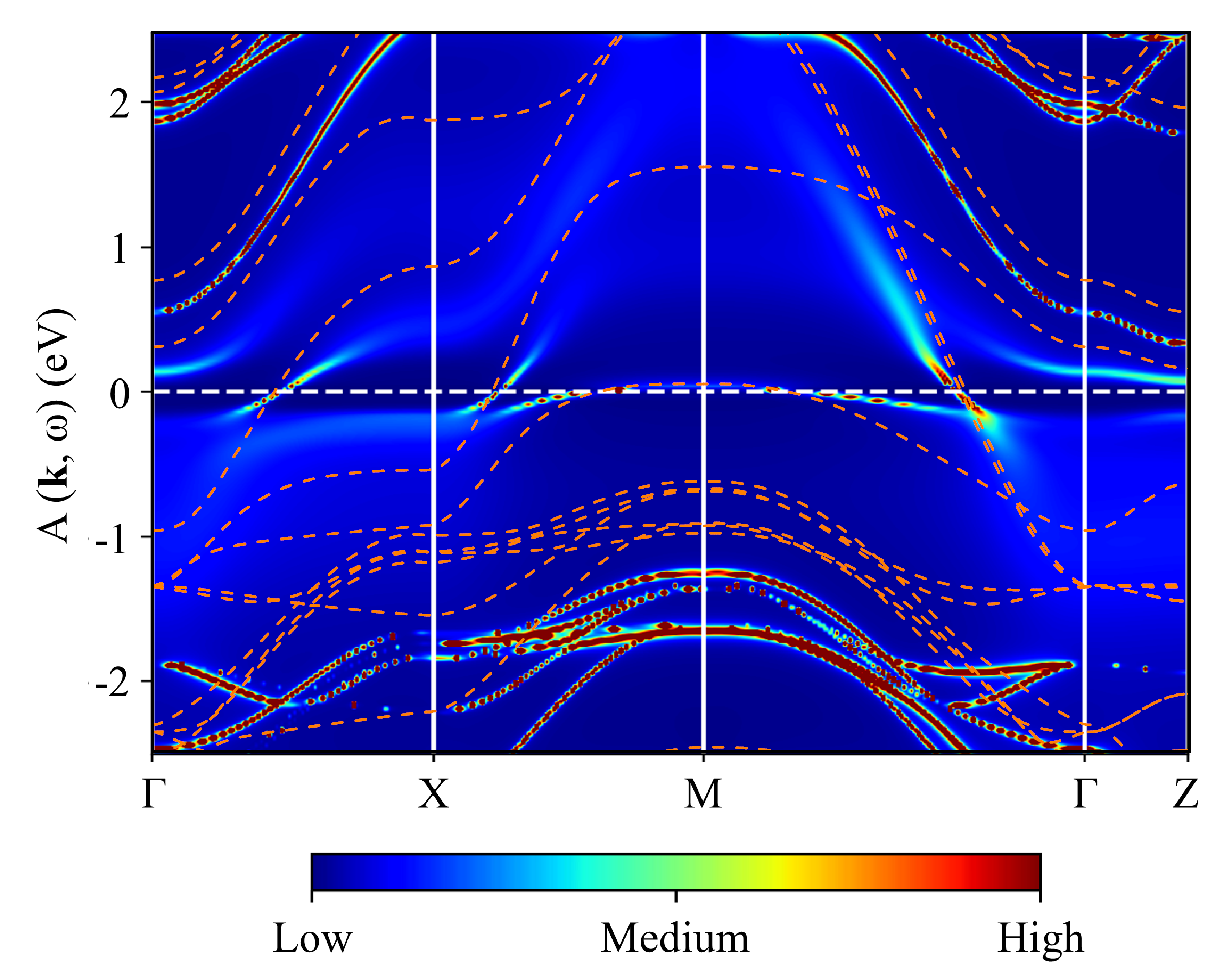}
\caption{Momentum-resolved spectral functions $A(\bf{k},\omega)$ obtained by DFT+DMFT calculations at $T =$ 80 K. The DFT band structures are plotted with orange dashed lines. The Fermi level is set to zero.}
\label{fig:band}
\end{figure}

\begin{figure*}[htb]
\centering
\includegraphics[width=17.6cm]{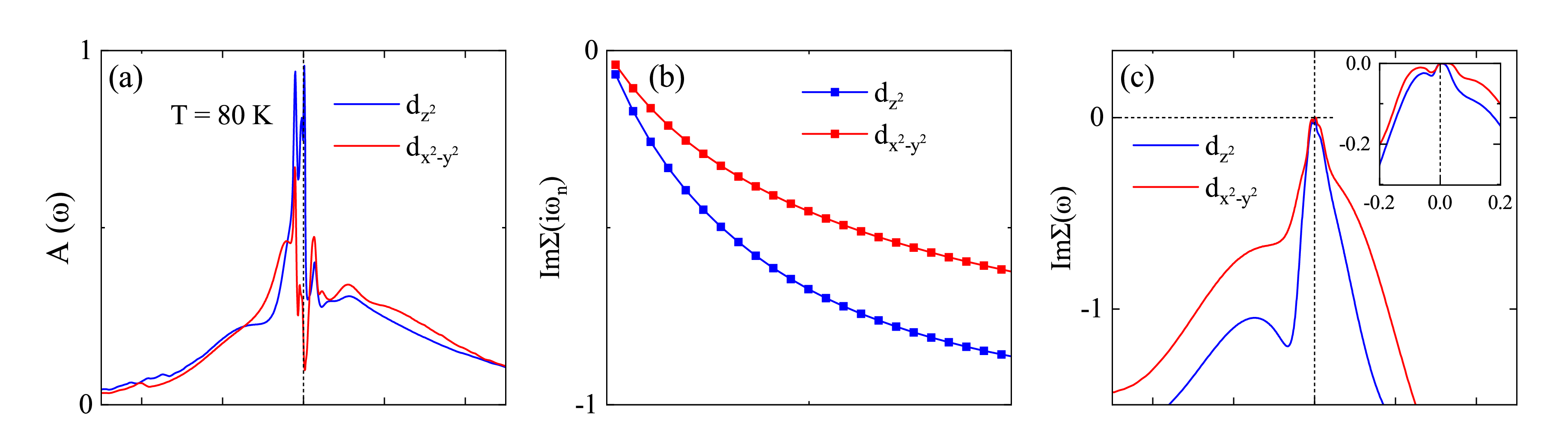}
\includegraphics[width=17.6cm]{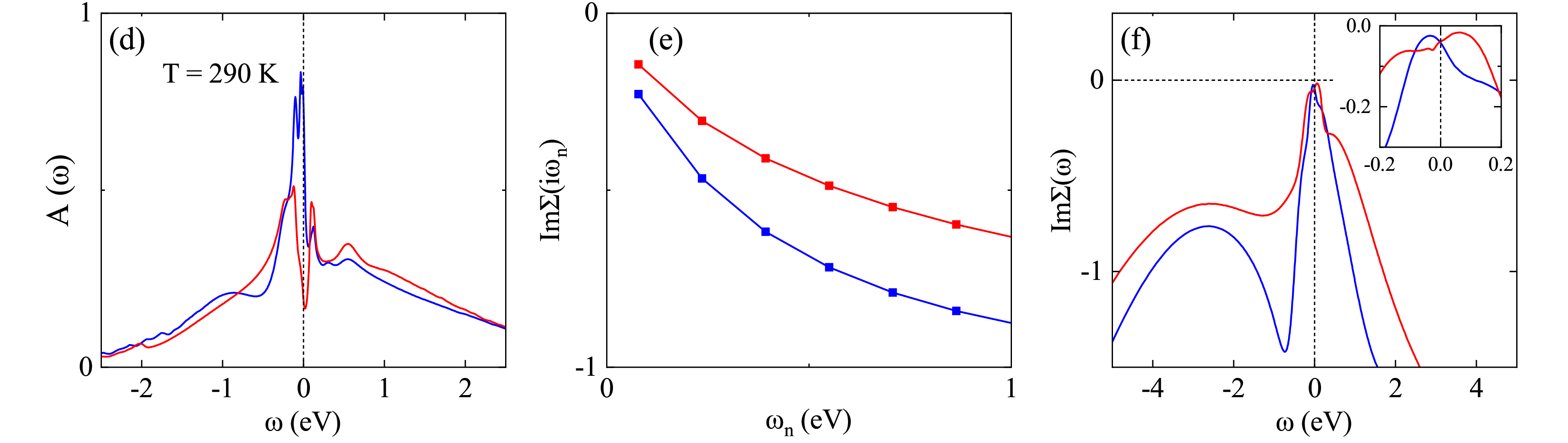}
\caption{Orbital-resolved spectral functions $A(\omega)$, the imaginary parts of self-energy functions at Matsubara axis Im$\Sigma (i\omega)$ and real axis Im$\Sigma (\omega)$ obtained by DFT+DMFT calculations for (a)-(c) at $T =$ 80 K and (d)-(f) at $T =$ 290 K. Inset in (c) and (f): the real axis Im$\Sigma (\omega)$ around $\omega =$ 0.}
\label{fig:2}
\end{figure*}

Figure~\ref{fig:band} shows the DFT band structures and momentum-resolved spectral functions $A (\bf{k},\omega)$ calculated by DFT+DMFT at $T =$ 80 K. Both DFT bands and DFT+DMFT spectral functions exhibit that three bands cross the Fermi level. Among them, Ni-3$d_{x^2-y^2}$ and 3$d_{z^2}$ orbitals contribute two electron pockets respectively around the $\Gamma$ and X points and Ni-3$d_{z^2}$ orbitals contribute a hole pocket at the M point. All these electronic bands are consistent with those in the previous reports. The strong inter-layer coupling between Ni-3$d_{z^2}$ and apical O-2$p_z$ orbitals leads to the flat bands at the M point split into a bonding state with lower energy and a higher energy antibonding state. Our DFT+DMFT calculations suggest that those bands around the Fermi level are strongly renormalized due to the electronic correlation. The bandwidth of Ni-3$d_{z^2}$ bonding state is compressed to 0.3 eV and the antibonding state moves to 0.7 eV above the Fermi level. In addition, we find that the antibonding state locating above the Fermi level at the M point is blurry. This is because a finite temperature makes an electronic incoherent-coherent crossover and a large proportion of electronic excitations are washed out. Similar behavior can be observed in those spectra below the Fermi level. When the temperature is higher, it is expected that the flat band at the M point will become incoherent. But those bands around the Fermi level already show good coherence at $T =$ 80 K.

\subsection{Orbital selective electronic correlation}
The strong band renormalization suggests electronic correlation in La$_3$Ni$_2$O$_7$. It is natural to wonder whether the origin of electronic correlation belongs to Mottness or Hundness. In order to study this problem, in Fig.~\ref{fig:2}, we show the orbital-resolved spectral functions $A(\omega)$, imaginary parts of the orbital-dependent self-energy at Matsubara axis Im$\Sigma (i\omega)$ and real axis Im$\Sigma (\omega)$. First of all, both 3$d_{z^2}$ and 3$d_{x^2 - y^2}$ orbitals appear around the Fermi level, which indicates the multi-obital metallic physics for La$_3$Ni$_2$O$_7$. For the 3$d_{z^2}$ orbitals, two narrow quasiparticle coherence peaks appear within a small energy window around the Fermi level, while the 3$d_{x^2-y^2}$ orbitals show two broader but lower peaks around the Fermi level, as shown in Fig.~\ref{fig:2}(a) and (d). As the temperature decreases from $T =$ 290 K to $T =$ 80 K, those coherence peaks for both orbitals become sharper, which is consistent to the trend towards forming a coherent Fermi liquid at lower temperature~\cite{Deng-NC10}.

\begin{table}[htb]
\begin{center}
\renewcommand\arraystretch{1.5}
\caption{The mass enhancement $m^*/m$ and the local occupancy number $N_{e_g}$ for the ${e_g}$ orbitals of Ni obtained by DFT+DMFT calculations under different temperatures.}
\label{tab1}
\begin{tabular*}{8.6cm}{@{\extracolsep{\fill}} cccccc}
\hline\hline

\multicolumn{2}{c} {Temperature (K)}                 & 80    & 150    & 220   & 290   \\

\hline

\multirow{2}*{$m^*$/$m$} & 3$d_{z^2}$                 & 3.394      & 3.019       & 2.736       & 2.516       \\
                                         & 3$d_{x^2 - y^2}$         & 2.529      & 2.333       & 2.157       & 2.015       \\

\hline

\multirow{2}*{$N_{e_g}$}       & 3$d_{z^2}$                  & 1.140      & 1.139       & 1.139       & 1.139       \\
                                         & 3$d_{x^2 - y^2}$          & 1.056      & 1.056      & 1.056       & 1.056       \\

\hline\hline
\end{tabular*}
\end{center}
\end{table}

Then, we show the imaginary parts of self-energy functions at the Matsubara axis Im$\Sigma (i\omega)$ in Fig.~\ref{fig:2}(b) and (e). Orbital selective electronic correlation is observed at both 80 and 290 K. Furthermore, we also show the imaginary parts of real axis self-energy Im$\Sigma(\omega)$ in Fig.~\ref{fig:2}(c) and (f) to roughly describe a likely phase transition in La$_3$Ni$_2$O$_7$. At high temperature if the system is in a frozen moment phase, electrons suffering scattering from the moments is expected so that Im$\Sigma(\omega = 0)$ is equal to a finite value $\Gamma$, which is corresponding to the intercept Im$\Sigma (i\omega \rightarrow 0) = \Gamma\text{sgn}(\omega_n)$~\cite{Werner-PRL101,Stewart-RMP73}. And the Fermi liquid behavior may be expected at lower temperatures with a linearly vanishing self-energy (Im$\Sigma (i\omega) \sim \omega$) at small $\omega$. Our calculated imaginary parts of self-energy functions at real axis Im$\Sigma (\omega)$ show such a trend. An intercept Im$\Sigma(\omega = 0) = \Gamma$ is observed at $T =$ 290 K but disappears at $T =$ 80 K. This finding suggests that the crossover from the frozen moment phase~\cite{Deng-NC10,Werner-PRL101} to the Fermi liquid phase may happen around 80 K.

In order to further characterize the orbital selective electronic correlation in La$_3$Ni$_2$O$_7$. Then, we list the orbital mass enhancement in Table~\ref{tab1}, which is a key factor to characterize the bands renormalization, as well as the electronic correlation. The mass enhancement is defined as:
\begin{equation}
{m^*}/m = {Z^{-1}} = 1 - \frac{\partial \mbox{Im} \Sigma (i\omega)}{\partial \omega} \bigg|_{\omega \rightarrow 0} .
\label{eq:massenhancement}
\end{equation}
$Z$ is the renormalization factor, which can be estimated from self-energy functions at the Matsubara axis. We find a mass enhancement difference of 2.5 and 2.0 for the 3$d_{z^2}$ and 3$d_{x^2 - y^2}$ orbitals at $T =$ 290 K, respectively. Such a difference becomes more significant when temperature is lower, with a stronger mass enhancement of 3.4 for 3$d_{z^2}$ and weaker mass enhancement of 2.5 for 3$d_{x^2 - y^2}$ at $T =$ 80 K. The mass enhancement confirms the strong renormalization of band structures in Fig.~\ref{fig:band}. And the difference of mass enhancement suggests the orbital selective electronic correlation in La$_3$Ni$_2$O$_7$. Furthermore, we show the local occupancy number $N_{e_g}$ for the $e_g$ orbitals of Ni. According to the nominal electron count, a Ni$^{2.5+}$ cation with 3$d^{7.5}$ occupancy is expected in La$_3$Ni$_2$O$_7$. Considering the crystal field of the Ni-O octahedra, $t_{2g}$ orbitals are fully filled, the occupancy number $N_{e_g}$ is thus about 1.5. However, as shown by our DFT+DMFT calculation results, a larger occupancy of 2.1-2.2 is found in the $e_g$ orbitals. At other different temperature conditions, the occupancy of 3$d_{z^2}$ and 3$d_{x^2 - y^2}$ orbitals is close to be half filled. Both orbital selective electronic correlation and near half filled occupancy in 3$d_{z^2}$ and 3$d_{x^2 - y^2}$ orbitals indicate possible characteristic multi-orbital Hund's metal behaviors in La$_3$Ni$_2$O$_7$.

\subsection{Hundness from high-spin states}
Here, we survey the local spin multiplets to demonstrate the Hund correlation in La$_3$Ni$_2$O$_7$. In those strong correlated materials that exhibit Hundness, the high-spin states are usually favored~\cite{Stadler-AP405,Yin-NM10}. In Table~\ref{tab2}, we list the probability $P_{\Gamma}$ of all local spin multiplets for given atomic configurations $\ket{\Gamma}$, which are labeled by some good quantum numbers, such as total spin $S_{\text{z}}$ and total occupancy $N_\Gamma$. For the $e_g$ orbitals of Ni, up to four electrons are allowed to be filled in. Among these atomic configurations under $T =$ 290 K, the high-spin state $S_{\text{z}} =$ 1 only exists when total occupancy $N_{e_g} =$ 2, which accounts for approximately 33.28$\%$, while the low-spin state with $N_{e_g} =$ 2, $S_{\text{z}} =$ 0 possesses a smaller percentage of 23.62$\%$. Such a difference indicates the existence of Hund interorbital correlation in La$_3$Ni$_2$O$_7$. Additionally, we find the charge fluctuations with 0.39$\%$, 12.36$\%$, 28.06$\%$, and 2.29$\%$ multiplets in $N =$ 0, 1, 3 and 4, respectively. Thus, the averaged occupancy of the $e_g$ orbitals $\braket{N_{e_g}} \equiv \Sigma_{\Gamma} N_{\Gamma}P_{\Gamma}$ is about 2.19, which is consistent with $N_{e_g}$ in Table~\ref{tab1} and larger than the nominal occupancy of 1.5 with Ni$^{2.5+}$. And the atomic configurations under $T =$ 80 K also show that the high-spin state possesses a larger percentage. All these results suggest electronic correlation of Hundness in La$_3$Ni$_2$O$_7$.

\begin{table}[htb]
\begin{center}
\small
\renewcommand\arraystretch{1.5}
\caption{Weights of the Ni ${e_g}$ orbital local multiplets obtained by DFT+DMFT calculations at temperature $T$ of 80 and 290 K, respectively.}
\label{tab2}
\begin{tabular*}{8.6cm}{@{\extracolsep{\fill}} ccccccc}
\hline\hline

$N_{\Gamma}$   & 0   & 1   & 2   & 2   & 3   & 4   \\

$S_z$  & 0    & 1/2   & 0   & 1   & 1/2   & 0  \\

\hline

$P_\Gamma$ (80 K)   & 0.40\%   & 12.41\%   & 23.85\%   & 32.94\%   & 28.10\%  & 2.30\%  \\

$P_\Gamma$ (290 K)    & 0.39\%   & 12.36\%   & 23.62\%   & 33.28\%   & 28.06\%  & 2.29\%  \\

\hline\hline
\end{tabular*}
\end{center}
\end{table}

\subsection{Hundness from imaginary-time correlation}
\begin{figure}[b]
\centering
\includegraphics[width=8.6cm]{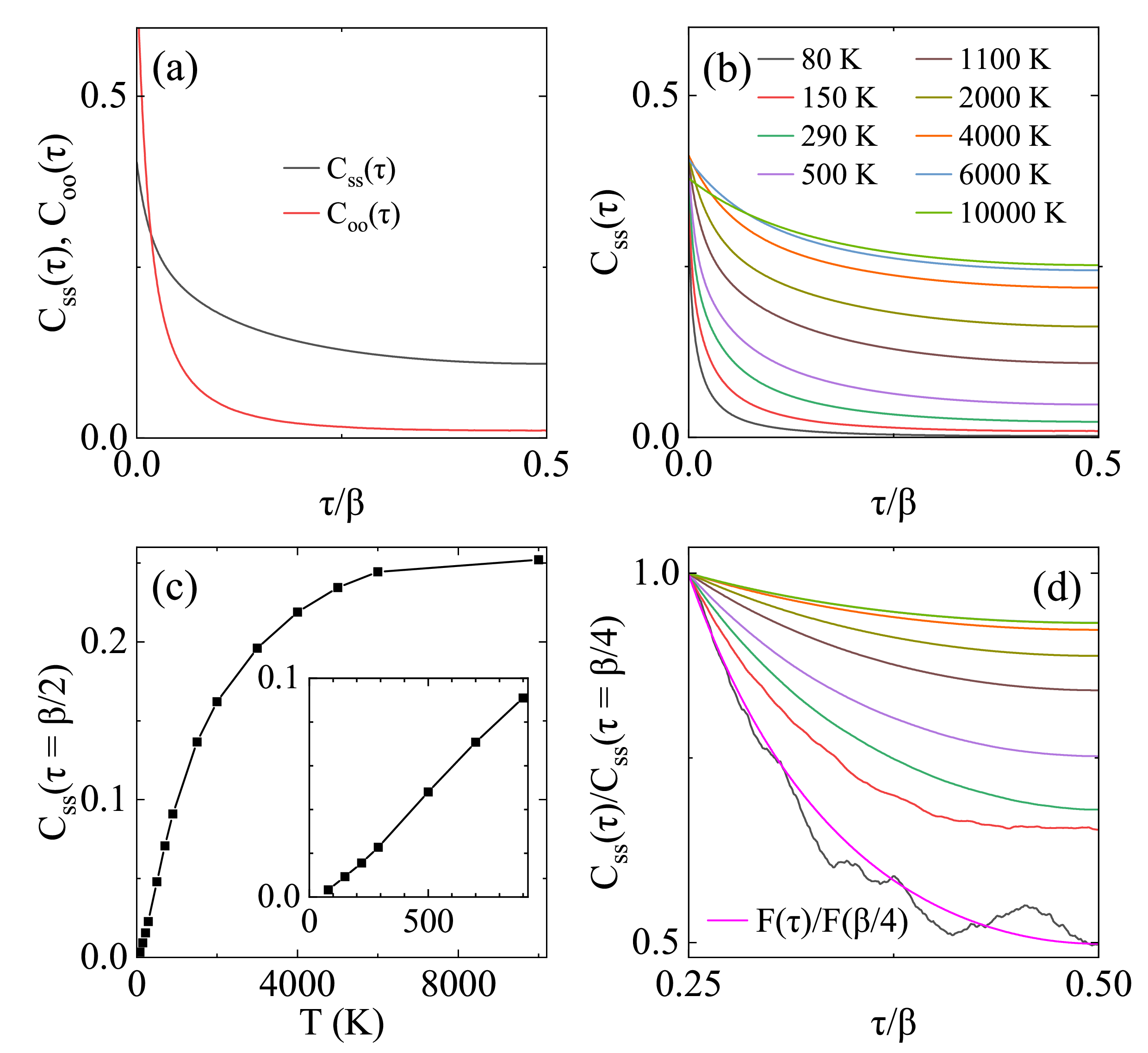}
\caption{ (a) Imaginary time spin-spin correlation functions $C_{ss}(\tau)$ and orbital-orbital correlation functions $C_{oo}(\tau)$ obtained by DFT+DMFT calculations at $T =$ 1100 K. (b) Imaginary time spin-spin correlation functions $C_{ss}(\tau)$ at different temperatures. (c) $C_{ss}(\tau = \beta/2)$ plotted as function of temperature. (d) Calculated $C_{ss}(\tau)/C_{ss}(\tau = \beta/4)$ at different temperatures, where $F(t) = (T/\sin(\pi\tau/\beta))^2$.}
\label{fig:3}
\end{figure}
Then, we study the spin-spin correlation function $C_{ss}(\tau) = \braket{S_{z}(\tau)S_{z}(0)}$ and orbital-orbital correlation function $C_{oo}(\tau) = \braket{O(\tau)O(0)}$ obtained by DFT+DMFT calculations at different temperatures, where $S_{z}$ is the total spin, and $O$ means $n(d_{x^2-y^2}) - n(d_{z^2})$. In Fig.~\ref{fig:3}(a), firstly, we see that at a finite temperature of $T =$ 1100 K, $C_{ss}(\tau)$ decays much slower than $C_{oo}(\tau)$ with imaginary time, which indicates spin-orbital separation~\cite{Stadler-AP405,Huang-PRB102}.
In Fig.~\ref{fig:3}(b), we exhibit a series of calculated $C_{ss}(\tau)$ under different temperatures. A large finite value of $C_{ss}(\tau = \beta/2)$ is observed above $T =$ 1100 K, which suggests that the system is in spin frozen phase~\cite{Werner-PRL101,Stadler-AP405}. It should be noted that here we only want to simulate the behaviors of electronic system under high temperature scales, such an extremely high temperature is not actually applied to the real crystal system, but just applied to electrons like external field to detect the corresponding responses of the electrons. Both spin-orbital separation and spin frozen phase are key signatures of Hundness, which have been observed in some typical Hund metal such as Sr$_2$RuO$_4$~\cite{Deng-NC10}.
In Fig.~\ref{fig:3}(c), a $C_{ss}(\tau = \beta/2) \sim T$ scaling is found above $T =$ 290 K and a $C_{ss}(\tau = \beta/2) \sim T^2$ behavior is found at low temperatures. Such scaling behaviors are also found in a three-orbital model study by Werner $\emph{et al}.$~\cite{Werner-PRL101}. $C_{ss}(\tau = \beta/2)$ $\sim T$ scaling is expected to emerge near the phase transition point from a frozen moment phase to a Fermi liquid phase~\cite{Werner-PRL101}. Besides, at a low temperature Fermi liquid phase, $C_{ss}(\tau) \sim F(\tau) = (T/\sin(\pi\tau{T}))^2$ for imaginary times $\tau$ sufficiently far from either $\tau = 0$ or $\tau = \beta$ and $C_{ss}(\tau) \sim T^2$ at $\tau = \beta/2$.
In Fig.~\ref{fig:4}(d), $C_{ss}(\tau = \beta/2)$ at $T =$ 80 K shows such a behavior, indicating that the system is in a Fermi liquid phase at low temperatures.

\subsection{Hundness from susceptibilities}
\begin{figure}[htb]
\centering
\includegraphics[width=8.6cm]{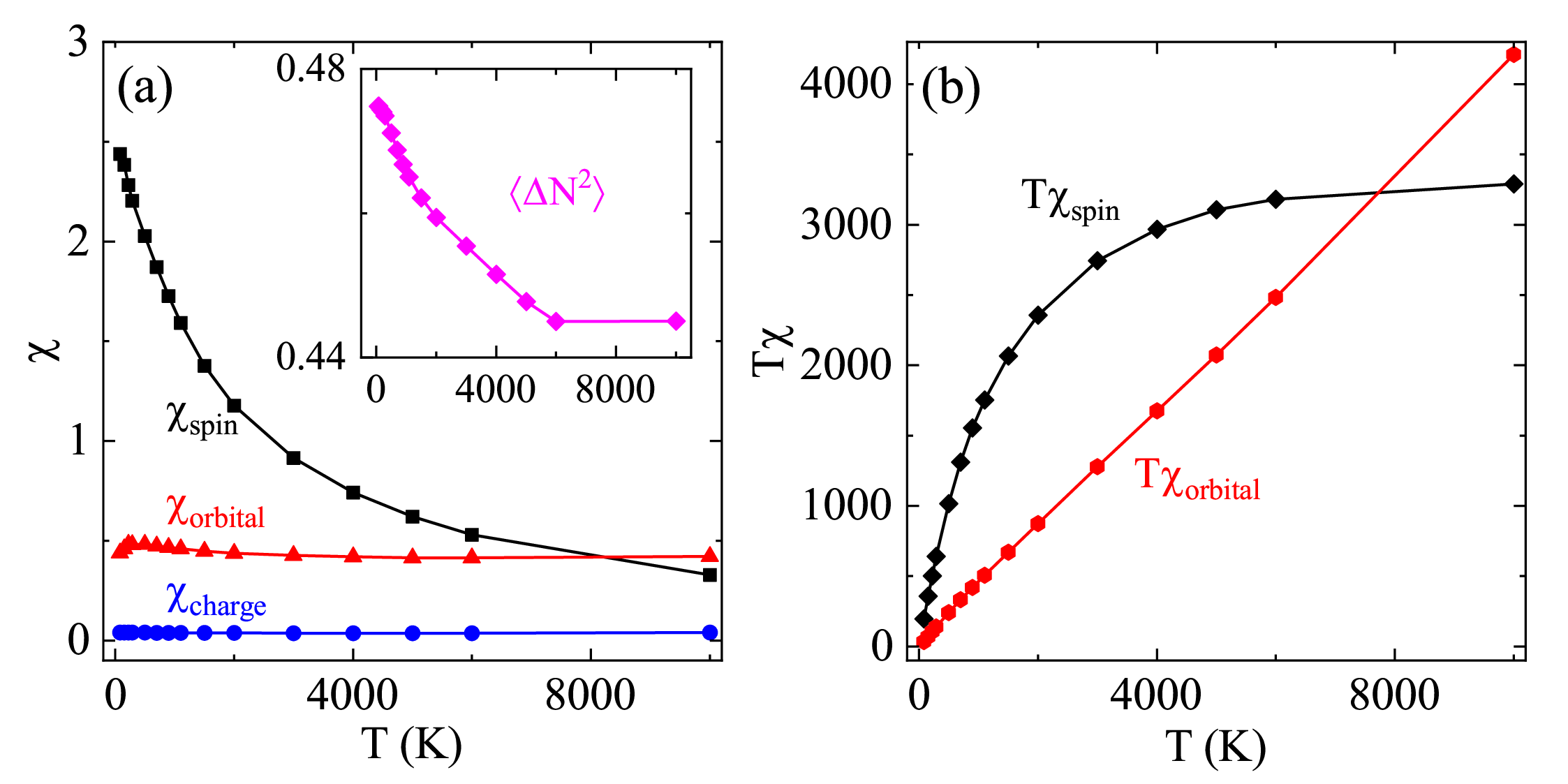}
\caption{(a) Static local susceptibilities $\chi$ for spin, orbital, and charge obtained by DFT+DMFT calculations under different finite temperatures. (b) $T\chi_{\text{spin}}$ and $T\chi_{\text{orbital}}$ plotted as functions of temperature. Inset in (a): local charge fluctuation $\braket{\Delta N^2}$. }
\label{fig:4}
\end{figure}
To further elucidate the origin of electronic correlation in La$_3$Ni$_2$O$_7$. We then study the static local susceptibilities under different temperatures. First of all, four temperature scales are identified, which characterize the onset and completion of screening of the orbital and spin degrees of freedom as the temperature is lowered. $T_{\text{orbital}}^{\text{onset}}$ and $T_{\text{spin}}^{\text{onset}}$ mark the onset temperatures of the screening for orbital and spin degrees of freedom, respectively. $T_{\text{orbital}}^{\text{cmp}}$ and $T_{\text{spin}}^{\text{cmp}}$ are defined as the completion temperatures of the screening~\cite{Deng-NC10}. When the temperature is lower than $T^{\text{onset}}$, the static local orbital and spin susceptibilities $\chi_{\text{orbital}}$ and $\chi_{\text{spin}}$ deviate from the Curie-Weiss behavior $\chi \sim \frac{1}{T}$. Below $T^{\text{cmp}}$ these susceptibilities become constant, which is known as the Pauli behavior. In a Mottness system, the screening of orbitals and spins begins almost simultaneously, as well as the completion ($T_{\text{orbital}}^{\text{onset}}$ $\sim$ $T_{\text{spin}}^{\text{onset}}$). In contrast, a Hundness system possesses a smaller value of $U$ and a larger value of $J_H$, which causes the screening of orbitals has started before the spin screening at a much higher temperature ($T_{\text{orbital}}^{\text{onset}}$ $\gg$ $T_{\text{spin}}^{\text{onset}}$). This is the so-called spin-orbital separation. Usually, such a behavior in a Hundness system features a broad temperature window than in a Mottness system. And in such a spin-orbital separation phase, the completely screened orbital degrees of freedom shows a Pauli behavior while $\chi_{\text{spin}}$ deviates from a Curie behavior. Meanwhile, a non-Fermi liquid behavior is observed in this phase till both the spin and orbital screenings complete. This is a key signature to identify the Hund electronic correlation in correlated materials.

In Fig.~\ref{fig:4}, we show the static local susceptibilities of spin, orbital, and charge plotted as functions of temperature. The $\chi_{\text{spin}}$, $\chi_{\text{orbital}}$, and $\chi_{\text{charge}}$ are defined as:
\begin{align}
\chi_{\mathrm{spin}} &= \int_{0}^{\beta} \braket{S_{z}(\tau)S_{z}(0)}d\tau, \\
\chi_{\mathrm{orbital}} &= \int_{0}^{\beta} \braket{O(\tau)O(0)}d\tau - \beta\braket{O}^2, \\
\chi_{\mathrm{charge}} &= \int_{0}^{\beta} \braket{N(\tau)N(0)}d\tau - \beta\braket{N}^2,
\end{align}
where $O = n(d_{x^2 - y^2}) - n(d_{z^2})$ is the orbital occupancy difference, $N$ is the total occupancy of the two $e_g$ orbitals. From a very high temperature of $T =$ 10000 K, we find that $\chi_{\text{orbital}}$ are almost unchanged when lowering the temperature. The unchanged $\chi_{\text{orbital}}$ strongly suggests that the screening of orbitals has completed at an extremely higher temperature than $T =$ 10000 K. However,  $\chi_{\text{spin}}$ and $\chi_{\text{orbital}}$ show obviously different behaviors when changing temperature. As the temperature decreases, $\chi_{\text{spin}}$ generally shows an increasing trend, while $\chi_{\text{orbital}}$ remains constant. We then display the $T\chi_{\text{spin}}$ and $T\chi_{\text{orbital}}$ curves versus temperature in Fig.~\ref{fig:4}(b). We see that the variation of $T\chi_{\text{orbital}}$ keeps a linear behavior, which deviates from the Curie-Weiss law. In contrast,  $T\chi_{\text{spin}}$ shows a Cuire behavior above $T =$ 5000 K, which is approximately constant. When the temperature is lower than $T =$ 5000 K, the Curie behavior is broken, indicating the onset of spin screening. And a broad non-Fermi liquid phase appears, which is corresponding to the spin-orbital separation. Within the process of spin screening, $C_{ss}(\tau)$ decays as the temperature decreases [Fig.~\ref{fig:3}(b)]. And a tendency towards constant for $\chi_{\text{spin}}$ begins to be observed at $T =$ 80 K, which suggests that the spin screening completes. This finding is consistent to our previous results, for example, the zero intercept of Im$\Sigma (\omega = 0)$ and the scaling behavior of $C_{oo}(\tau = \beta/2) \sim T^2$ at $T =$ 80 K, showing the low temperature Fermi liquid phase.

\subsection{Absence of Mottness}
\begin{figure}[hb]
\centering
\includegraphics[width=8.6cm]{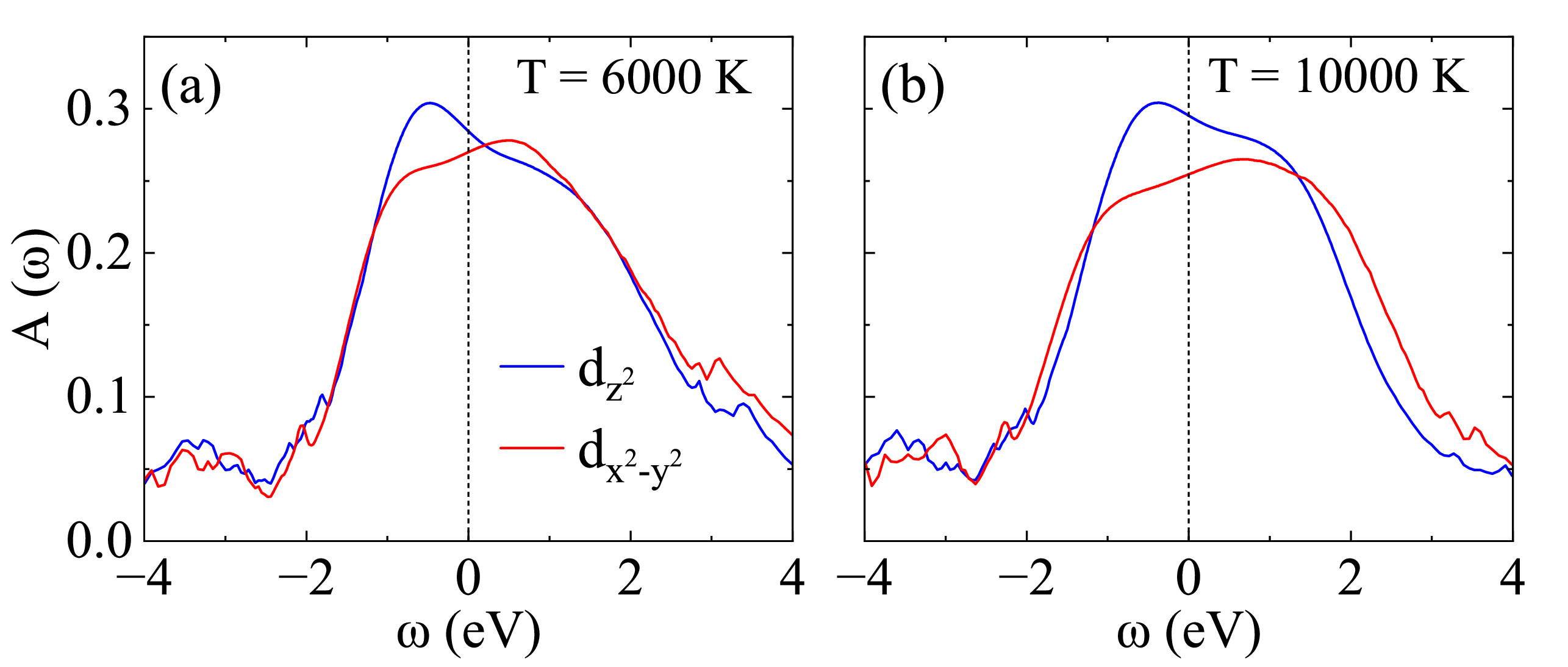}
\caption{Orbital-resolved spectral functions $A(\omega)$ obtained by DFT+DMFT calculations for (a) $T =$ 6000 K and (b) $T =$ 10000 K. The blue and red lines represent the 3$d_{z^2}$ and 3$d_{x^2-y^2}$ orbitals of Ni, respectively.  }
\label{fig:5}
\end{figure}
In order to check the Mott electronic correlation in La$_3$Ni$_2$O$_7$, we exhibit the orbital-resolved spectral functions $A(\omega)$ at larger temperatures, where the spin screening has not started. For instance, in V$_2$O$_3$, which is an archetypal Mott correlated metal~\cite{Deng-NC10}. Before the onset of spin screening, a pseudogap feature is observed in V$_2$O$_3$. But no pseudogap is observed in La$_3$Ni$_2$O$_7$ under $T =$ 6000 K and 10000 K (Fig.~\ref{fig:5}). In addition, we also find that both the local charge fluctuation $\braket{\Delta N^2} = \braket{N^2} - \braket{N}^2$ and $\chi_{\text{charge}}$ are almost unchange when the temperature is lower [Fig.\ref{fig:4}(a)], while in a Mottness system $\braket{\Delta N^2}$ should have an obvious temperature dependence, manifesting delocalization of electrons~\cite{Deng-NC10}. All these findings suggest that the electronic correlation in La$_3$Ni$_2$O$_7$ is attributed to Hund's physics.

\section{CONCLUSION}
In summary, we have studied the electronic structures and correlation of La$_3$Ni$_2$O$_7$ via DFT+DMFT calculations. The  momentum-resolved spectral functions $A(\bf{k},\omega)$ show that La$_3$Ni$_2$O$_7$ is a multi-orbital metal. The 3$d_{z^2}$ and 3$d_{x^2 - y^2}$ orbitals of Ni contribute those bands across the Fermi level. The electronic correlation leads to a strong band renormalization, where the band width of bonding state band at the M point is reduced remarkably. The orbital-resolved spectral functions $A(\omega)$ show sharp quasiparticle coherence peaks around the Fermi level. The mass enhancement $m^*/m$ derived from the self-energy functions at the Matsubara axis reveals the orbital selective electronic correlation. The electronic correlation of the 3$d_{z^2}$ orbital is stronger than that of the 3$d_{x^2 - y^2}$ orbital. At $T =$ 80 K, the mass enhancements of the 3$d_{z^2}$ and 3$d_{x^2 - y^2}$ orbitals are about 3.4 and 2.5, respectively. A higher percentage of high-spin state suggests the existence of inter-orbital Hund electronic correlation within the $e_g$ orbitals of Ni. Moreover, the calculated spin-spin and orbital-orbital correlation functions show spin-orbital separation and frozen moments at high temperatures. The scaling behavior of the spin-spin correlation functions indicates that La$_3$Ni$_2$O$_7$ is in a spin-frozen phase at high temperatures ($T > 290$ K) or in a Fermi liquid phase at low temperatures. The calculated static local spin, orbital, and charge susceptibilities under high temperatures also show signatures of Hundness. All these results show that La$_3$Ni$_2$O$_7$ is a multi-orbital Hund's metal. Our DFT+DMFT calculations uncover the Hund electronic correlation in La$_3$Ni$_2$O$_7$ and provide a new perspective to understand the origin of electronic correlation in correlated materials.

\begin{acknowledgments}
This work was supported by the National Natural Science Foundation of China (under Grants No. 11934020). L. Huang was also supported by the CAEP Foundation (under Grant No. CX20210033).
Computational resources were provided by the Physical Laboratory of High Performance Computing at Renmin University of China.
\end{acknowledgments}

\end{document}